\RequirePackage{fix-cm}
\documentclass[smallextended]{svjour3}       
\smartqed  
\usepackage{graphicx}
\usepackage{algorithmic}
\usepackage{amssymb,amsmath}
\usepackage{caption}
\usepackage{subcaption}
\usepackage{setspace} 
\usepackage{multirow}
\usepackage{booktabs}
\usepackage{multicol}
\usepackage{hyperref}

\usepackage{caption}
\usepackage{color}

\usepackage{listings}
\definecolor{codegreen}{rgb}{0,0.6,0}
\definecolor{codegray}{rgb}{0.5,0.5,0.5}
\definecolor{codepurple}{rgb}{0.58,0,0.82}
\definecolor{backcolour}{rgb}{0.95,0.95,0.92}

\lstdefinestyle{mystyle}{
    backgroundcolor=\color{backcolour},   
    commentstyle=\color{codegreen},
    keywordstyle=\color{magenta},
    numberstyle=\tiny\color{codegray},
    stringstyle=\color{codepurple},
    basicstyle=\ttfamily\footnotesize,
    breakatwhitespace=false,         
    breaklines=true,                 
    captionpos=b,                    
    keepspaces=true,                 
    numbers=left,                    
    numbersep=5pt,                  
    showspaces=false,                
    showstringspaces=false,
    showtabs=false,                  
    tabsize=2
}

\usepackage[linesnumbered,ruled]{algorithm2e}

\usepackage[top=1in,bottom=1in,left=1in,right=1in]{geometry}

\begin{document}

\title{\textit{SpeedyIBL}: A Comprehensive, Precise, and Fast Implementation of Instance-Based Learning Theory

}

\titlerunning{Speedy IBL}        

\author{Thuy Ngoc Nguyen \and Duy Nhat Phan \and \\ Cleotilde Gonzalez*\thanks{*Corresponding author: Cleotide Gonzalez} }


\institute{Thuy Ngoc Nguyen         \and Duy Nhat Phan \and Cleotilde Gonzalez \at
              5000 Forbes Ave., Porter Hall 223-G, Pittsburgh, 15213 PA, USA \\
              \email{ngocnt@cmu.edu, dnphan@andrew.cmu.edu, coty@cmu.edu}           
}

\date{Received: date / Accepted: date}

\maketitle

\begin{abstract}
Instance-Based Learning Theory (IBLT) is a comprehensive account of how humans make decisions from experience during dynamic tasks. Since it was first proposed almost two decades ago, multiple computational models have been constructed based on IBLT (i.e., IBL models). These models have been demonstrated to be very successful in explaining and predicting human decisions in multiple decision making contexts. However, as IBLT has evolved, the initial description of the theory has become less precise, and it is unclear how its demonstration can be expanded to more complex, dynamic, and multi-agent environments. This paper presents an updated version of the current theoretical components of IBLT in a comprehensive and precise form. It also provides an advanced implementation of the full set of theoretical mechanisms, \textit{SpeedyIBL}, to unlock the capabilities of IBLT to handle a diverse taxonomy of individual and multi-agent decision making problems. SpeedyIBL addresses a practical computational issue in past implementations of IBL models, the curse of exponential growth, that emerges from memory-based tabular computations. When more observations accumulate over time, there is an exponential growth of the memory of instances that leads directly to an exponential slow down of the computational time. Thus, SpeedyIBL leverages parallel computation with vectorization to speed up the execution time of IBL models. We evaluate the robustness of SpeedyIBL over an existing implementation of IBLT in decision games of increased complexity. The results not only demonstrate  the applicability of IBLT through a wide range of decision making tasks, but also highlight the improvement of SpeedyIBL over its prior implementation as the complexity of decision features and number of agents increase. The library is open sourced for the use of the broad research community.


\end{abstract}

\section{Introduction}
A cognitive theory is a general postulation of mechanisms and processes that are globally applicable to families of tasks and types of activities rather than being dependent on a particular task. Cognitive models are very specific representations of part or all aspects of a cognitive theory that apply to a particular task or activity \cite{gonzalez2017decision}. Specifically, normative and descriptive theories of choice often rely on utility theory \cite{Savage1954,morgenstern1953theory} or aim at describing the psychological impact of perceptions of probability and value on choice \cite{kahneman1979prospect,tversky1992advances}. In contrast, models of decisions from experience (DfE) are often dynamic computational representations of sequential choices that are distributed over time and space and that are made under uncertainty \cite{gonzalez2017dynamic}.

Cognitive models of DfE can be used to simulate the interaction of theoretical cognitive processes with the environment, representing a particular task. These models can make predictions regarding how human choices are made in such tasks. These predictions are often compared to data collected from human participants in the same tasks using interactive tools. The explicit comparison of cognitive models' predictions to human actual behavior is a common research approach in the cognitive sciences and in particular in the study of decision making \cite{gonzalez2017decision}.
Cognitive models are dynamic and adaptable computational representations of the cognitive structures and mechanisms involved in decision making tasks such as DfE tasks under conditions of partial knowledge and uncertainty. Moreover, cognitive models are generative, in the sense that they actually make decisions in similar ways like humans do, based on their own experience, rather than being data-driven and requiring large training sets. In this regard, cognitive models differ from purely statistical approaches, such as Machine Learning models, that are often capable of evaluating stable, long-term sequential dependencies from existing data but fail to account for the dynamics of human cognition and human adaptation to novel situations.

There are many models of DfE as evidenced by past modeling competitions \cite{erev2010choice,erev2017anomalies}. Most of these models often make broadly disparate assumptions regarding the cognitive processes by which humans make decisions \cite{erev2010choice}. For example, the models submitted to these competitions are often applicable to a particular task or choice paradigm rather than presenting an integrated view of how the dynamic choice process from experience is performed by humans. Associative learning models are a class of models of DfE that conceptualize choice as a \textit{learning} process that stores behavior-outcome relationships and are contingent on the environment \cite{hertwig2015}. A common example of this type of models is reinforcement learning (RL) \cite{sutton2018reinforcement}, and the association between DfE and RL is becoming more explicit in the literature \cite{konstantinidis2020memory,speekenbrink2015uncertainty}. Generally speaking, these kinds of models rely on learning from reinforcement and the contingencies of the environment as in the Skinnerian tradition \cite{skinner2014contingencies,sutton1995theory}. These models have shown to be successful at representing human learning over time based on feedback. 

In contrast to many of the associative learning models, \textit{Instance-Based Learning} (IBL) models rely on a single dynamic decision theory: Instance-Based Learning Theory (IBLT) \cite{GONZALEZ03}. IBLT emerged from the need to explain the process of dynamic decision making, where a sequence of interdependent decisions are made sequentially, over time. IBLT provides a single general algorithm and mathematical formulations of memory retrieval that rely on the well-known ACT-R cognitive architecture \cite{ANDERSON14}. The theory proposes a representation of decisions in the form of \textit{instances}, which are triplets involving state, action, and utilities. In general, states are a representation of the features of the situation of the environment in a task, actions are decisions an agent makes in such states, and utilities are the expectations the agent generates or the outcomes the agent receives from performing such actions. The theory also provides a process of retrieval of past instances based on their similarity to a current decision situation, and the generation of accumulated value (expectation from experience) based on a mechanism called \textit{Blending}, which is a function of the payoffs experienced and the probability of retrieving those instances from memory \cite{LEBIERE99,LEJARRAGA12,GONZALEZ11}. 

Initially, IBLT was demonstrated in a highly complex, dynamic decision making task representing the complex process of dynamic allocation of limited resources over time and under time constraints in a ``water purification plant'' \cite{GONZALEZ03}. Since its inception, many models have been developed based on IBLT, demonstrating human DfE in a large diversity of contexts and domains, from simple and abstract binary choice dynamics \cite{GONZALEZ11,LEJARRAGA12}, to highly specialized tasks such as cyber defense \cite{aggarwal2020exploratory,cranford2020toward} and anti-phishing detection \cite{cranford2019modeling}. Also, IBL models have been created to account for dyadic and group effects, where each individual in a group is represented by an IBL agent \cite{gonzalez2015cognitive}. More recently, this IBL algorithm has been applied to multi-state gridworld tasks \cite{NGUYEN20,NGUYEN2020ICCM,Ngoc2021} in which the agents execute a sequence of actions with delayed feedback. The recent applications of IBLT have led to significantly more complex and realistic tasks, where multi-dimensional state-action-utility representations are required, where extended training is common, where real-time interactivity between models and humans is needed to solve such tasks \cite{Ngoc2021}. 

With the increased use of IBLT in generating models on tasks of greater complexity and in multiple domains, it has become clear that the initial, two-decade old conceptualization of IBLT needs to be updated. As IBLT has evolved, the initial description of the theory has become less precise, given that no formal implementation of the IBLT process was provided. Thus, a comprehensive description of the current state of the theory along with a concrete implementation of the whole IBL process is essential. Moreover, it is important to demonstrate the full capability and generality of IBLT in a single manuscript, that explains and illustrates how models of multiple and diverse decision tasks can be constructed based on the same theory to generate predictions regarding DfE and learning across a wide range of decision making tasks. With that, the major goal of this paper is to provide an updated view of the theoretical components of IBLT in a comprehensive and precise form.  We also provide an open source, efficient implementation of the full set of mechanisms of IBLT and demonstrate how such implementation can handle a diverse taxonomy of individual and multi-agent decision making tasks.

In the process of generating IBL models for more complex tasks that require real-time interactivity between models and humans, we have confronted a practical computational problem, the \textit{curse of exponential growth} \cite{bellman1957dynamic,kuo2005lifting}. The curse of exponential growth is a common problem in models that rely on the accumulation of data over time and on computation of approximate value functions represented as arrays and tables, such as RL models \cite{sutton2018reinforcement}. 
As summarized in a recent overview of the challenges in multi-agent RL models, even advanced deep reinforcement learning techniques with many successes in Atari, Go, and Starcraft games~\cite{mnih2013playing,silver2016mastering,vinyals2019alphastar} suffer severely from the increase in the dimensions of the state-action space, particularly as the number of agents increases \cite{wong2021multiagent}. The problem becomes even more complex under nonstationary environments and under uncertainty, where information is incomplete. Dynamic conditions significantly increase the diversity and number of states as it is needed for every dynamic decision making task \cite{gonzalez2017dynamic}. Thus, this paper also addresses the critical question of how IBL models can tackle the curse of exponential growth of memory.

In summary, we present three main contributions. First, an updated view of IBLT provides a comprehensive and precise view of the current theoretical components of the theory, offering a concrete generic algorithm with a formal implementation of the general process of IBLT. Second, we demonstrate the applicability of IBLT across a taxonomy of decision-making tasks varying in the number of agents, the number of actions, the number of decision options and states, and the type of delayed feedback. Third, we provide a new, open source library, \textit{SpeedyIBL}, that can handle the curse of exponential growth. SpeedyIBL allows users to create multiple IBL agents relying on IBLT with fast processing and response time while maintaining the decision characteristics of IBL models. We demonstrate how SpeedyIBL is increasingly beneficial (compared to existing implementations, PyIBL~\cite{MorrisonGonzalez}) as the dimensions of the representation, the number of agents and their interactions increase. Through simulation experiments, we demonstrate how IBL models are able to provide predictions across a taxonomy of decision-making tasks with escalating complexity, and how SpeedyIBL is increasingly more efficient than PyIBL~\cite{MorrisonGonzalez} as the dimensions of task complexity increase.

\section{Instance based Learning Theory}
\begin{figure}[!htbp]
    \centering
    \includegraphics[width=0.8\textwidth]{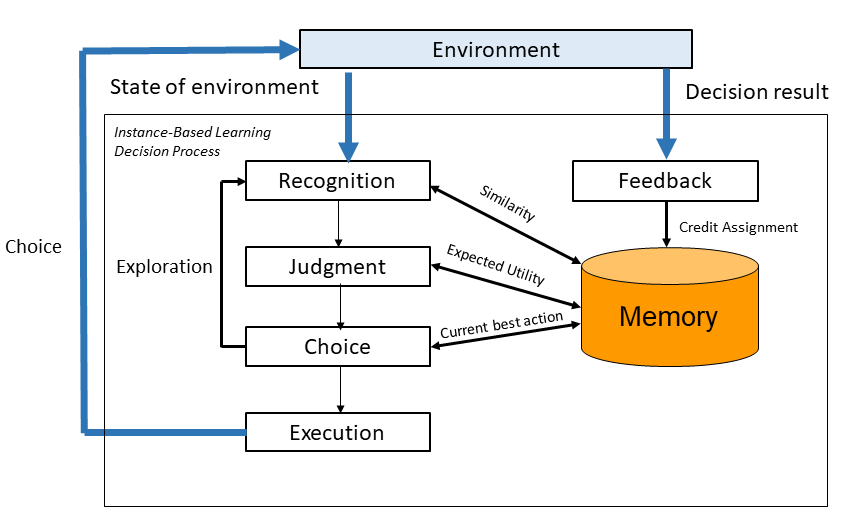}
    \caption{IBLT algorithm from \cite{GONZALEZ03}}
    \label{fig:iblt}
\end{figure}

An updated view of the general decision process proposed in IBLT is illustrated in Figure \ref{fig:iblt}, and the current mechanisms of IBLT are made mathematically concrete in Algorithm \ref{alg:IBL} \cite{GONZALEZ03}.

The process starts with the observation of the environmental state, and the determination of whether there are past experiences in memory (i.e., instances) that are similar to the current environmental state (i.e., Recognition). Whether there are similar past instances will determine the process used to generate the expected utility of a decision alternative (i.e., Judgment). If there are past experiences that are similar to the current environmental state, the expected utility of such an alternative is calculated via a process of \textit{Blending} past instances from memory; but if there are no similar past instances, then the theory suggests that a heuristic is used to generate the expected utility, instead. After Judgment, the option with the highest expected utility is maintained in memory and a decision is made as to whether to stop the exploration of additional alternatives and execute the current best decision (i.e., Choice) or to continue exploring new alternatives (i.e., exploration Loop). When the exploration process ends, the choice that has the highest expected utility is executed, which changes the environment (i.e., Execution Loop). The loop from Recognition to Execution continues over time, and the result from a decision may be observed from the environment (i.e., Feedback) immediately or with delay from the execution of a choice. Such decision result (e.g., a reward) is used to update the utility of past instances in memory through a \textit{credit assignment} mechanism.

In IBLT, an ``instance'' is a memory unit that results from the potential alternatives evaluated. These memory representations consist of three elements which are constructed over time: a situation state $s$ which is composed of a set of features $f$; a decision or action $a$ taken corresponding to an alternative in state $s$; and an expected utility or experienced outcome $x$ of the action taken in a state.

Each instance in memory has an \textit{Activation} value, which represents how readily available that information is in memory, and it is determined by the similarity to past situations, recency, frequency, and noise according to the Activation equation in ACT-R \cite{ANDERSON14}. Activation of an instance is used to determine the probability of retrieval of an instance from memory which is a function of its activation relative to the activation of all instances corresponding the same state in memory. The expected utility of a choice option is calculated by blending past outcomes. This blending mechanism for choice has its origins in a more general blending formulation \cite{LEBIERE99}, but a simplification of this mechanism is often used in models with discrete choice options, defined as the sum of all past experienced outcomes weighted by their probability of retrieval \cite{GONZALEZ11,LEJARRAGA12}. This formulation of blending represents the general idea of an \textit{expected value} in decision making, where the probability is a cognitive probability, a function of the activation equation in ACT-R. Algorithm \ref{alg:IBL} provides a formal representation of the general IBL process.

\begin{algorithm}[ht!]
\caption{Pseudo Code of Instance-based Learning process} 
\label{alg:IBL} 
\SetKwInput{KwInput}{Input}
\SetKwFor{ExecutionLoop}{Execution Loop}{}{end}
\SetKwFor{ExplorationLoop}{Exploration Loop}{do}{end}
\DontPrintSemicolon

\KwInput{default utility $x_0$, a memory dictionary $\mathcal M= \{\}$, global counter $t = 1$, step limit $L$, a flag $delayed$ to indicate whether feedback is delayed.}
\Repeat{task stopping condition}{  
Initialize a counter (i.e., step) $l=0$ and observe state $s_l$

\While{$s_l$ is not terminal and $l<L$}{
   \ExecutionLoop{}{
  \ExplorationLoop{$a\in A$}{
   Compute activation values $\Lambda_{i(s_l^i,a)t}$ of instances $((s_l^i,a), x_{i(s_l^i,a)t},T_{i(s_l^i,a)t})$ by \eqref{eq:activation}\;\label{algo:activation}
   
Compute retrieval probabilities $P_{i(s_l^i,a)t}$ by \eqref{eq:retrieval_prob}
\;\label{algo:prob}
Compute blended values $V_{(s_l,a)t}$ corresponding to $(s_l,a)$ by \eqref{eq:blended_value}\; \label{algo:blended}
}
Choose an action $a_l\in\arg\max_{a\in A}V_{(s_l,a)t}$
}
   Take action $a_l$, move to state $s_{l+1}$, observe $s_{l+1}$, and receive outcome $x_{l+1}$\;
   Store $t$ into instance corresponding to selecting $(s_l,a_l)$ and achieving outcome $x_{l+1}$ in $\mathcal M$\;\label{algo:store}
   If $delayed$ is true, update outcomes using a \textit{credit assignment} mechanism\;\label{algo:delay}
   $l \leftarrow l+1$  and $t \leftarrow t+1$\;
   }
 }
\end{algorithm}

Concretely, for an agent, an option $k=(s,a)$ is defined by taking action $a$ after observing state $s$.
At time $t$, assume that there are $n_{kt}$ different considered instances $(k_i,x_{ik_it})$ for $i = 1,...,n_{kt}$, associated with $k$. Each instance $i$ in memory has an \textit{Activation} value, which represents how readily available that information is in memory and expressed as follows~\cite{ANDERSON14}: 

 \begin{equation}\label{eq:activation}
 \begin{array}{l}
     \Lambda_{ik_it} = \ln{\left(\sum\limits_{t' \in T_{ik_it} }(t-t')^{-d}\right)} + \alpha\sum\limits_{j}Sim_j(f^k_j,f^{k_i}_j)  + \sigma\ln{\frac{1-\xi_{ik_it}}{\xi_{ik_it}}},
\end{array}
 \end{equation}

where $d$, $\alpha$, and $\sigma$ are the decay, mismatch penalty, and noise parameters, respectively, and $T_{ik_it}\subset \{0,...,t-1\}$ is the set of the previous timestamps in which the instance $i$ was observed, $f_j^k$ is the $j$-th attribute of the state $s$, and $Sim_j$ is a similarity function associated with the $j$-th attribute. The second term is a partial matching process
reflecting the similarity between the current state $s$ and the state of the option $k_i$. The rightmost term represents a noise for capturing individual variation in activation, and $\xi_{ik_it}$ is a random number drawn from a uniform distribution $U(0, 1)$ at each timestep and for each instance and option.

Activation of an instance $i$ is used to determine the probability of retrieval of an instance from memory.
The probability of an instance $i$ is defined by a soft-max function as follows
 \begin{equation} \label{eq:retrieval_prob}
     P_{ik_it} = \frac{e^{\Lambda_{ik_it}/\tau}}{\sum_{j = 1}^{n_{kt}}e^{\Lambda_{jk_jt}/\tau}},
 \end{equation}
where $\tau$ is the Boltzmann constant (i.e., the ``temperature") in the Boltzmann distribution.
For simplicity, $\tau$ is often defined as a function of the same $\sigma$ used in the activation equation $\tau= \sigma\sqrt{2}$.

The expected utility of option $k$ is calculated based on \textit{Blending} as specified in choice tasks~\cite{LEJARRAGA12,GONZALEZ11}:
 \begin{equation} \label{eq:blended_value}
     V_{kt} = \sum_{i=1}^{n_{kt}}P_{ik_it}x_{i k_i t}.
 \end{equation}
The choice rule is to select the option that corresponds to the maximum blended value. In particular, at the $l$-th step of an episode, the agent selects the option $(s_l,a_l)$ with
\begin{equation}
  a_l = \arg\max_{a\in A}  V_{(s_l,a),t}
\end{equation}

The flag $delayed$ on line \ref{algo:delay} of Algorithm \ref{alg:IBL} is true when the agent knows the real outcome after making a sequence of decision without feedback. In such case, the agent updates outcomes by using one of the credit assignment mechanisms~\cite{Nguyen21}. 
It is worth noting that when the flag $delayed$ is true depends on a specific task. For instance, $delayed$ can be set to true when the agent reaches the terminal state, or when the agent receives a positive reward. 

\section{SpeedyIBL Implementation} \label{sec:speedyibl}
From the IBL algorithm \ref{alg:IBL}, we observe that its computational cost revolves around the computations on lines \ref{algo:activation} (Eq. \ref{eq:activation}),  \ref{algo:prob} (Eq. \ref{eq:retrieval_prob}), \ref{algo:blended} (Eq. \ref{eq:blended_value}), and the storage of instances with their associated time stamps on line \ref{algo:store}.  
Clearly, when the number of states and action variables (dimensions) grow, or the number of IBL agent objects increases, the execution of steps \ref{algo:activation} to \ref{eq:blended_value}) in algorithm \ref{alg:IBL} will directly increase the execution time. The ``speedy'' version of IBL (i.e., SpeedyIBL) is a library focused on dealing with these computations more efficiently. 

SpeedyIBL algorithm is the same as that in Algorithm \ref{alg:IBL}. The innovation is in the Mathematics. Equations \ref{eq:activation}, \ref{eq:retrieval_prob} and \ref{eq:blended_value} are replaced with Equations \ref{eq:activation2}, \ref{eq:retrieval_prob2} and \ref{eq:blended_value2}, respectively (as explained below). Our idea is to take advantage of \textit{vectorization}, which typically refers to the process of applying a single instruction to a set of values (vector) in parallel, instead of executing a single instruction on a single value at a time. In general, this idea can be implemented in any programming language. We particularly implemented these in Python, since that is how PyIBL is implemented \cite{MorrisonGonzalez}.

Technically, the memory in an IBL model is stored by using a dictionary $\mathcal M$ that, at time $t$, represented as follows:
\begin{equation}
    \mathcal M = \biggl\{k_i: \{x_{ik_it}: T_{ik_it}, ...\}, ...\biggr\},
\end{equation}
where $(k_i,x_{ik_it},T_{ik_it})$ is an instance $i$ that corresponds to selecting option $k_i$ and achieving outcome $x_{ik_it}$ with $T_{ik_it}$ being the set of the previous timestamps in which the instance $i$ is observed. 

To vectorize the codes, we convert $T_{ik_it}$ to a \texttt{NumPy}\footnote{\url{https://numpy.org/doc/stable/}} array \cite{harris2020array} on which we can use standard mathematical functions with built-in \texttt{Numpy} functions for fast operations on entire arrays of data without having to write loops. 

After the conversion, we consider $T_{ik_it}$ as a \texttt{NumPy} array. In addition, since we may use a common similarity function for several attributes, we assume that $f$ is partitioned into $J$ non-overlapping groups $f_{[1]},...,f_{[J]}$ with respect to the distinct similarity functions $Sim_1,...,Sim_J$, i.e., $f_{[j]}$ contains attributes that use the same similarity function $Sim_j$. We denote $S(f^k,f^{k_i})$ the second term of \eqref{eq:activation} computed by:

\begin{algorithmic}
\STATE set $S(f^k,f^{k_i})$ = 0
\FOR{$j=1$ to $J$}
 \STATE{$S(f^k,f^{k_i})\ += \texttt{sum}((Sim_j(f_{[j]}^k,f_{[j]}^{k_i}))$}
\ENDFOR
\end{algorithmic}
Hence, the activation value (see Equation \ref{eq:activation}) can be fast and efficiently computed as follows:
\begin{equation}\label{eq:activation2}
    \Lambda_{ik_it} = \texttt{math.log}(\texttt{sum}(\texttt{pow}(t-T_{ik_it},-d))) + \alpha*S(f^k,f^{k_i}) + \sigma*\texttt{math.log}((1-\xi_{ik_it})/\xi_{ik_it}).
\end{equation}
With the vectorization, the operation such as $\texttt{pow}$ can be performed on multiple elements of the array at once, rather than looping through and executing them one at a time. 
Similarly, the retrieval probability (see Equation \ref{eq:retrieval_prob}) is now computed by:
\begin{equation} \label{eq:retrieval_prob2}
    P_{kt} := [P_{1k_1t},...,P_{n_{kt}k_{n_{kt}}t}] = v/\texttt{sum}(v),
\end{equation}
where $v = \texttt{math.exp}([\Lambda_{1k_1t},...,\Lambda_{n_{kt}k_{n_{kt}}t}]/\tau)$. The blended value (see Equation \ref{eq:blended_value}) is now computed by:
\begin{equation}\label{eq:blended_value2}
    V_{kt} = \texttt{sum}(x_{kt}*P_{kt}),
\end{equation}
where $x_{kt}: = [x_{1k_1t},...,x_{n_{kt}k_{n_{kt}}t}]$ is a \texttt{NumPy} array that contains all the outcomes associated with the option $k$.

   
    

\section{Experiments: demonstration of the general applicability of IBLT}
To demonstrate the applicability of IBLT through a wide range of decision tasks as well as to assess the efficiency of SpeedyIBL, we compare SpeedyIBL performance against a regular implementation of the IBL algorithm (Algorithm \ref{alg:IBL}) in Python (PyIBL~\cite{MorrisonGonzalez}), in six different tasks that were selected to represent different dimensions of complexity in dynamic decision making tasks~\cite{gonzalez2005use}.

\subsection{A Taxonomy of Individual and Multi-Agent Decision-Making Tasks}
Generally, computational cognitive science has taken advantage of the availability of large amounts of behavioral data to advance the ``explanation'' of cognitive processes involved in various
types of tasks, notably, decision making (\cite{griffiths2015manifesto}). These models often make excellent predictions of human choices in a particular task. However, for the advancement of cognitive science, it is generally important not to simply make accurate predictions in a specific task but to also provide general explanations and understanding of how and why people behave the way they do across tasks.

The development of computational cognitive models that are based on cognitive theories are expected to provide prediction power without a heavy reliance on data \cite{hofman2021integrating}. IBLT is a general postulation of mechanisms and processes that are globally applicable to families of dynamic decision tasks, rather than being dependent on the requirements of a particular task. In this section we present a taxonomy of decision making tasks that IBLT can address.

Table \ref{tab:task} provides an overview of six dimensions to vary in six different decision making tasks: (1) number of agents, (2) number of actions, (3) complexity of the states, (4) number of choice options (i.e., alternatives), (5) similarity across states, and (6) feedback delays. The table also presents six tasks that were selected to illustrate how IBLT can handle these dimensions. Although we selected these six specific tasks to illustrate the generality of IBLT, it is important to note that the theory is applicable to any diversity of tasks within these dimensions. For example IBLT can handle any number of agents, actions, and other task complexities.

\begin{table}[ht]
\centering
\begin{tabular}{lcccccc} 
\toprule 
\multirow{2}{*}{Task} &Num. of &Num. of &Num. of &Num. of &Similarity &Delayed\\
&Agents &Actions &States &Options &Judgments &Feedback\\
\hline 
Binary choice & 1 & 2 & 1 & 2 & No & No\\

Insider attack game & 1 & 6 & 4 & 24 & Yes & Yes\\

Minimap & 1 & 4 & $\approx 10^{41}$ & $\approx 4\times 10^{41}$ & No & Yes\\

Ms.Pac-Man & 1 & 9 &  $\approx 10^{347}$ & $\approx 9\times 10^{347}$ & No & Yes\\
Fireman & 2 & 4 & $\approx 10^{15}$ & $\approx 4\times 10^{15}$ & No & Yes\\
Cooperative navigation & 3 & 4 & $\approx 10^7$ & $ \approx 4\times 10^7$ & No & Yes \\
\bottomrule
\end{tabular}
    \caption{Taxonomy of Decision Making dimensions, and the illustration of six decision making tasks}
    \label{tab:task}
\end{table}

In terms of the number of agents, we selected four single agent tasks, one task with two agents, and one task with three agents. The tasks selected for demonstration can have between two to nine potential actions, the number of states and choice options also vary from just a few to a significant large number. We also include one task that requires of similarity judgments across states (i.e., partial matching in equations \ref{eq:activation} and \ref{eq:activation2}) and five tasks that do not use similarity judgments. Finally, we include one task with immediate feedback and five tasks that involve feedback delays.

We describe each of the tasks below, starting from the simplest task (repeated \textit{Binary choice}), and moving up in the level of task complexity. The binary choice task has only one state and two options; the \textit{Insider attack} task is a two-stage game in which players choose one of six targets after observing their features to advance. We then scale up to a larger number of states and actions in significantly more complex tasks. A \textit{Minimap} task representing a search and rescue mission and \textit{Ms. Pac-Man} tasks have a larger number of discrete state-action variables. Next, we scale up to two multi-agent tasks: the \textit{Fireman} task has two agents and four actions, and a \textit{Cooperative Navigation} task in which three agents navigate and cooperate to accomplish a goal. The number of agents increases the memory computation, since each of those agents adds their own variables to the joint state-action space. Based on these dimensions of increasing complexity, we expect that SpeedyIBL's benefits over PyIBL will be larger with increasing complexity of the task.


\subsubsection{Binary choice}\label{binarychoice}
In each trial, the agent is required to choose one of two options: Option A or Option B. A numerical outcome drawn from a distribution after the selection, is the immediate feedback of the task. This is a well-studied problem in the literature of risky choice task~\cite{Hertwig2004}, where individuals make decisions under uncertainty. Unknown to the agent is that the options A and B are assigned to draw the outcome from a predefined distribution. One option is safe and it yields a fixed medium outcome (i.e., $3$) every time it is chosen. The other option is risky, and it yields a high outcome ($4$) with some probability $0.8$, and a low outcome ($0$) with the complementary probability $0.2$. 

An IBL model of this task has been created and reported in various past studies, including \cite{GONZALEZ11,LEJARRAGA12}. 
Here, we conducted the simulations of 1000 runs of 100 trials. We also run the experiment with 5000 trials to more clearly highlight the difference between PyIBL and SpeedyIBL. The default utility $x_0$ was set to $4.4$. For each option $s$, where $s$ is either A or B, we consider all the generated instances taking the form of $(s,x)$, where $x$ is an outcome. 
The performance is determined by the average proportion of the maximum reward expectation choice (PMax).

\begin{figure}[!htbp]
    \centering
    \includegraphics[width=0.6\textwidth]{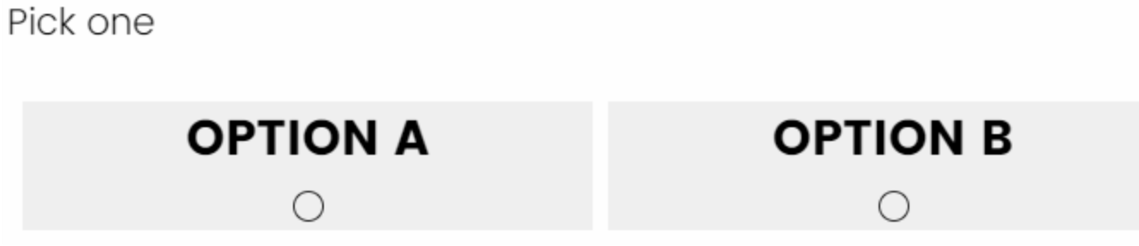}
    \caption{Binary choice}
    \label{fig:insider}
\end{figure}

\subsubsection{Insider attack game}
The insider attack game is an interactive task designed to study the effect of signaling algorithms in cyber deception experiments (e.g., \cite{Cranford18}). Figure~\ref{fig:insider} illustrates the interface of the task, including a representation of the agent (insider attacker) and the information of 6 computers.  Each of the six computers is ``protected'' with some probability (designed by a defense algorithm). Each computer displays the monitoring probability and potential outcomes and the information of the signal. When the agent selects one of the six computers, a signal is presented to the agent (based on the defense signaling strategy); which informs the agent whether the computer is monitored or not. The agent then makes a second decision after the signal: whether to continue or withdraw the attack on the pre-selected computer. If the agent attacks a computer that is monitored, the player loses points, but if the computer is not monitored, the agent wins points. The signals are, therefore, truthful or deceptive. If the agent withdraws the attack, it earns zero points.

\begin{figure}[!htbp]
    \centering
    \includegraphics[width=0.6\textwidth]{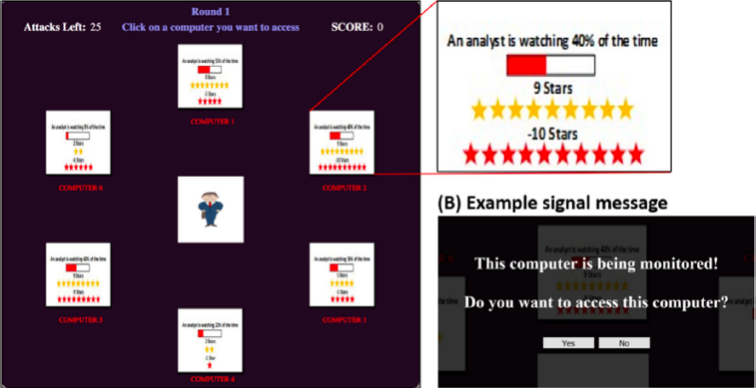}
    \caption{Insider Attack game}
    \label{fig:insider}
\end{figure}

In each trial, the agent must decide which of the 6 computers to attack, and whether to continue or withdraw the attack after receiving a signal. An IBL model of this task has been created and reported in past studies (e.g., \cite{cranford2019modeling,Cranford2021Towards}). We perform the simulations of 1000 runs of 100 episodes. For each option $(s,a)$, where the sate $s$ is the features of computers including reward, penalty and the probability that the computers is being monitored (see \cite{cranford2019modeling} for more details), and $a\in \{1,...,6\}$ is an index of computers, we consider all the generated instances taking the form of $(s',a,x)$ with $s'$ being a state and $x$ being an outcome. The performance is determined by the average collected reward.

\subsubsection{Search and rescue in Minimap}
The Minimap task is inspired by a search and rescue scenario, which involves an agent being placed in a building with multiple rooms and tasked with rescuing victims~\cite{Nguyen21b}. 
Victims have been scattered across the building and their injuries have different degrees of severity with some needing more urgent care than others. In particular, there are 34 victims grouped into two categories (24 green victims and 10 yellow victims). There are many obstacles (walls) placed in the path forcing the agent to look for alternative routes. The agent's goal is to rescue as many victims as possible. The task is simulated as a $93 \times 50$ grid of cells which represents one floor of this building. Each cell is either empty, an obstacle, or a victim. The agent can choose to move left, right, up, or down, and only move one cell at a time. 
\begin{figure}[!htbp]
	\centering
    \includegraphics[width=0.7\textwidth]{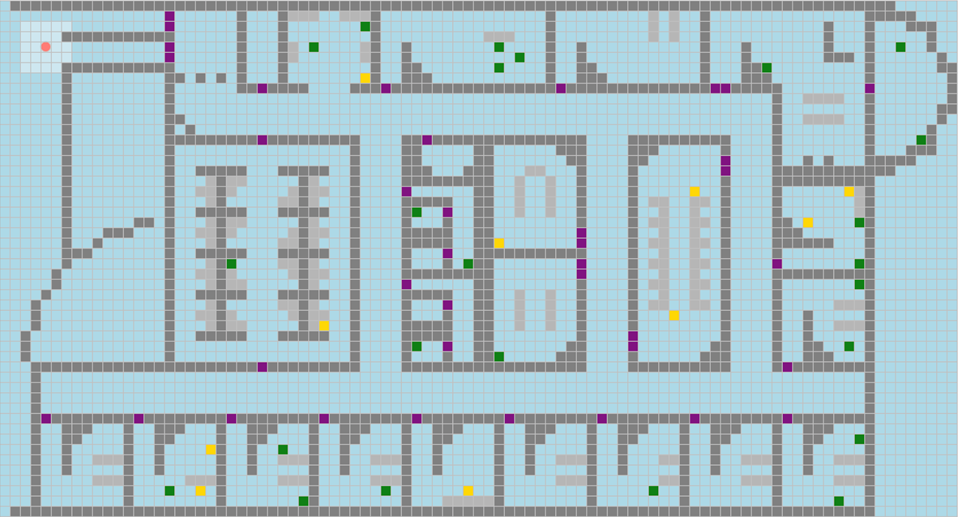} 
	\caption{Search and rescue map. The empty cells are white and the walls are black. The yellow and green cells represent the locations of the yellow and green victims respectively. The cell with the red color represents the start location of the agent.}
	
	\label{fig:minimap}
\end{figure}

The agent receives a reward of 0.75 and 0.25 for rescuing a yellow victim and a green victim, respectively. Moving into an obstacle or an empty cell is penalized by 0.05 or 0.01 accordingly. Since the agent might have to make a sequence of decisions to rescue a victim, we update the previous instances by a positive outcome that once the agent receives. 

An IBL model of this task has been created and reported in past studies \cite{Gulati2021Task}. Here we created the SpeedyIBL implementation of this model to perform the simulation of 100 runs of 100 episodes. An episode terminates when a $2500$-trial limit is reached or when the agent successfully rescues all the victims. After each episode, all rescued victims are placed back at the location where they were rescued from and the agent restarts from the pre-defined start position.


In this task, a state $s$ is represented by a gray-scale image (array) with the same map size.
We use the following pixel values to represent the entities in the map: $s[x][y]$ = 240 if the agent locates at the coordinate $(x,y)$, 150 if a yellow victim locates at the coordinate $(x,y)$, 200 if a green victim locates at the coordinate $(x,y)$, 100 if an obstacle locates at the coordinate $(x,y)$, and 0 otherwise. For each option $(s,a)$, where $s$ is a state and $a$ is an action, we consider all the generated instances taking the form of $(s,a,x)$ with $x$ being an outcome. The default utility was set to $0.1$. The flag $delayed$ is set to true if the agent rescues a victim, otherwise false. The performance is determined by the average collected reward. 

    


\subsubsection{Ms. Pac-Man}
\begin{figure}[!htbp]
    \centering
    \includegraphics[width=0.35\textwidth]{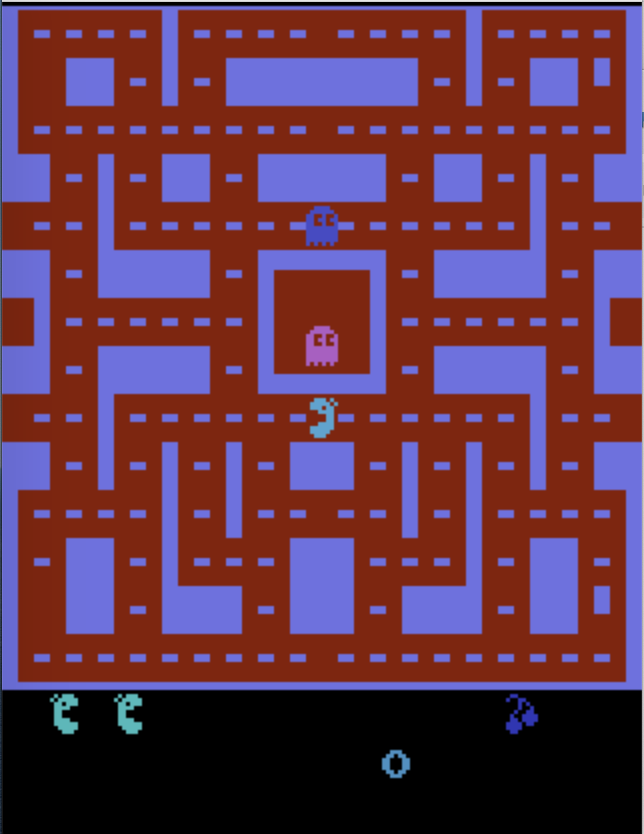}
    \caption{Mis.Pac-Man game}
    \label{fig:mis-pacman}
\end{figure}

The next task considered in the experiment is Ms. Pac-Man game, a benchmark for evaluating agents in machine learning, e.g. \cite{Hasselt2016Deep}. 
The agent maneuvers Pac-Man in a maze while Pac-Man eats the dots (see Fig.~\ref{fig:mis-pacman}). 

In this particular maze, there are 174 dots and each one is worth 10 points. A level is finished when all dots are eaten. To make things more difficult, there are also four ghosts in the maze who try to catch Pac-Man, and if they succeed, Pac-Man loses a life. Initially, she has three lives and gets an extra life after reaching $10,000$ points.
There are four power-up items in the corners of the maze, called power dots (worth
40 points). After Pac-Man eats a power dot, the ghosts turn blue for a short period, they slow down and try to escape from Pac-Man. During this time, Pac-Man is able to eat them, which is worth 200, 400, 800, and 1600 points, consecutively. The point values are reset to 200 each time another power dot is eaten, so the agent would want to eat all four ghosts per power dot. If a ghost is eaten, his remains hurry back to the center of the maze where the ghost is reborn. At certain intervals, a fruit appears near the center of the maze and remains there for a while. Eating this fruit is worth 100 points.

We use the MsPacman-v0 environment developed by Gym OpenAI\footnote{\url{https://gym.openai.com/envs/MsPacman-v0/}}, where a state is represented by a color image. Here, we developed an IBL model for this task and created the SpeedyIBL implementation of this model to perform the simulation of 100 runs of 100 episodes. An episode terminates when either a $2500$-step limit is reached or when Pac-Man successfully eats all the dots or loses three lives. Like in the Minimap task, for each option $(s,a)$, where $s$ is a state and $a$ is an action, we consider all the generated instances taking the form of $(s,a,x)$ with $x$ being an outcome.  The parameter $delayed$ is set to true if Pac-Man receives a positive reward, otherwise it is set to false. The performance is determined by the average collected reward.

\subsubsection{Fireman}
\begin{figure}[!htbp]
    \centering
    \includegraphics[width=0.35\textwidth]{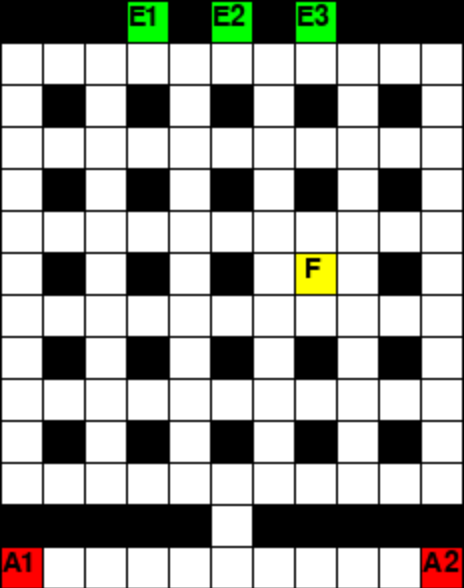}
    \caption{Fireman game}
    \label{fig:fireman}
\end{figure}

The Fireman task replicates the coordination in firefighting service wherein agents need to pick up matching items for extinguishing fire. This task was used for examining deep reinforcement learning agents \cite{Palmer2019Negative}. 
In the experiment, the task is simulated in a gridworld of size $11\times 14$, as illustrated in Fig. \ref{fig:fireman}. Two agents A1 and A2 located within the gridworld are tasked with locating an equipment pickup area and choosing one of the firefight items. Afterwards, they need to navigate and find the location of the fire (F) to extinguish it. The task is fully cooperative as both agents are required to extinguish one fire.
More importantly, the location of the fire is dynamic in every episode.

The agents receive the collective reward according to the match between their selected firefighting items, which is determined by the payoff matrix in Table \ref{tab:payoffmatrix}. The matrix is derived from a partial stochastic climbing game~\cite{MatignonLF12} that has a stochastic reward. If they both select the equipment E2, they get 14 points with the probability 0.5, and 0 otherwise. 
This Fireman task has both stochastic and dynamic properties. 

\begin{table}[!htbp]
    \centering
    \begin{tabular}{|l|l|l|l|l|} 
\hline
\multicolumn{2}{|c|}{} & \multicolumn{3}{|c|}{Agent 2}\\
\cline{3-5}
 \multicolumn{2}{|c|}{} & E1 & E2 & E3 \\
\hline
\multirow{3}{*}{\rotatebox[origin=c]{90}{Agent 1}} & E1 & 11 & -30 & 0\\
\cline{2-5}
& E2& -30 & 14/0 & 6\\
\cline{2-5}
& E3 & 0 & 0 & 5\\
\hline 
\end{tabular}
    \caption{Payoff matrix.}
    \label{tab:payoffmatrix}
\end{table}

Here we developed an IBL model for this task. We created the SpeedyIBL implementation of this model to perform the simulations of 100 runs of 100 episodes. An episode terminates when a $2500$-trial limit is reached or when the agents successfully extinguish the fire. After each episode, the fire is replaced in a random location and the agents restart from the pre-defined start positions.

Like in the search and rescue Minimap task, a state $s$ of the agent A1 (resp. A2) is represented by a gray-scale image with the same gridworld size using the following pixel values to represent the entities in the gridworld: $s[x][y]$ = 240 (resp. 200) if the agent A1 (resp. A2) locates at the coordinate $(x,y)$, 55 if the fire locates at the coordinate $(x,y)$, 40 if equipment E1 locates at the coordinate $(x,y)$, 50 if equipment E2 locates at the coordinate $(x,y)$, 60 if  equipment E3 locates at the coordinate $(x,y)$, 100 if an obstacle locates at the coordinate $(x,y)$, 0 otherwise.
Moreover, we assume that the agents cannot observe the relative positions of the other, and hence, their states do not include the pixel values of the other agent. For each option $(s,a)$, where $s$ is a state and $a$ is an action, we consider all the generated instances taking the form of $(s,a,x)$ with $x$ being an outcome. The flag $delayed$ is set to true if the agents finish the task, otherwise false. The performance is determined by the average collected reward.

\subsubsection{Cooperative Navigation} \label{navigation}
\begin{figure}[!htbp]
    \centering
    \includegraphics[width=0.35\textwidth]{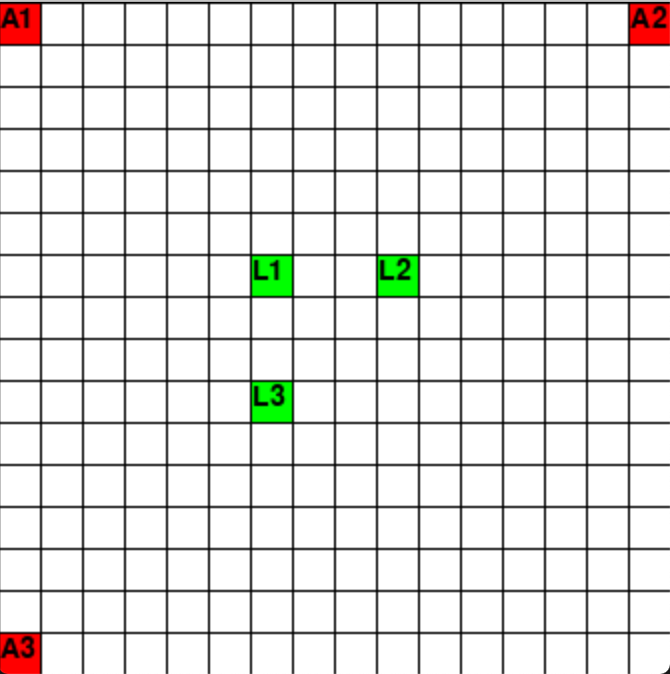}
    \caption{Cooperative navigation}
    \label{fig:navigation}
\end{figure}

In this task, three agents (A1, A2 and A3) must cooperate through physical actions to reach a set of three landmarks (L1, L2 and L3) shown in Fig.~\ref{fig:navigation}, see \cite{Lowe2017Multi}. The agents can observe the relative positions of other agents and landmarks, and are collectively rewarded based on the number of the landmarks that they cover. For instance, if all the agents cover only one landmark L2, they receive one point. By contrast, if they all can cover the three landmarks, they get the maximum of three points.
Simply put, the agents want to cover all landmarks, so they need to learn to coordinate the landmark they must cover. 

Here we developed an IBL model for this task. We created the SpeedyIBL implementation of this model to perform the simulations of 100 runs of 100 episodes. An episode terminates when a $2500$-trial limit is reached or when each of the agents covers one landmark. After each episode, the fire is replaced in a random location and the agents restart from the pre-defined start positions.

In this task, a state $s$ is also represented by a gray-scale image with the same gridworld size using the following pixel values to represent the entities in the environment: $s[x][y]$ = 240 if the agent A1 locates at the coordinate $(x,y)$, 200 if the agent A2 locates at the coordinate $(x,y)$, 150 if the agent A3 locates at the coordinate $(x,y)$, 40 if the landmark L1 locates at the coordinate $(x,y)$, 50 if the landmark L2 locates at the coordinate $(x,y)$, 60 if the landmark L3 locates at the coordinate $(x,y)$, 0 otherwise.
For each option $(s,a)$, where $s$ is a state and $a$ is an action, we consider all the generated instances taking the form of $(s,a,x)$ with $x$ being an outcome. The flag $delayed$ is set to true if the agents receive a positive reward, otherwise false. The performance is determined by the average collective reward.



\subsection{General Simulation Methods}
All the experiments are conducted on a PC AMD 3.00 GHz Ryzen 9 of 16GB RAM and 8 cores with Python 3.7.4 and Numpy 1.19.2. The detailed guideline on how to use the SpeedyIBL package is available at \url{https://github.com/DDM-Lab/SpeedyIBL} and the Appendix provides a detailed tutorial including installation of the SpeedyIBL library and examples on how to replicate our demonstrations in the tasks offered in this paper.

The parameter values configured in the IBL models with SpeedyIBL and PyIBL implementations were identical. 
In particular, we used the decay $d=0.5$ and noise $\sigma = 0.25$. The default utility values generally set to be higher than the maximum value obtained in the task, to create exploration as suggested in \cite{LEJARRAGA12} (see the task descriptions for specific values), and they were set the same for PyIBL and SpeedyIBL. 

For each of the six tasks, we compared the performance of PyIBL and SpeedyIBL implementations in terms of (i) running time measured in seconds and (ii) performance. The performance measure is identified within each task.

We conducted 1000 runs of the models and each run performed 100 episodes for the \textit{Binary choice} and \textit{Insider attack}. Given the running time required for PyIBL, we only ran 100 runs of 100 episodes for the remaining tasks. We note that an episode of the \textit{Binary choice} and \textit{Insider attack} tasks has one step (trial) while the remaining tasks have $2500$ steps within each episode. 

The credit assignment mechanisms in IBL are being studied in ~\cite{NGUYEN20}. In this paper we used an \textit{equal} credit assignment mechanism for all tasks. This mechanism updates the current outcome for all the actions that took place from the current state to the last state where the agent started or the flag $delayed$ was true.

\section{Results}
In this section, we present the results of the SpeedyIBL and PyIBL models across all the considered tasks. The comparison these packages is first provided in terms of the average running time and performance, and then in terms of their learning curves.

\subsection{Average Running time and Performance}
Table \ref{tab:runningtime} shows the overall average of computational time and Table \ref{tab:reward} the average performance across the runs and 100 episodes. The Ratio in Table \ref{tab:runningtime} indicates the speed improvement from running the model in SpeedyIBL over PyIBL.

\begin{table}[!htbp]
\resizebox{.99\textwidth}{!}{%
    \centering
    \begin{tabular}{lrlrlr} 
\toprule
Task &
PyIBL & & 
SpeedyIBL & &
Speed up\\
& time & & time & & Ratio \\
\midrule
Binary choice & 0.009 & & 0.008 & & 1.13\\
Insider Attack Game & 0.141 & & 0.065 & & 2.17 \\
Minimap &
21951.88 & ($\approx$ 365 mins $\approx$ 6 hours) &
78.40 & ($\approx$ 1.3 mins) &
279.00 \\
Ms.Pac-Man &
162372.58 & ($\approx$ 2706.2 mins $\approx$ 45 hours) &
111.98 & ($\approx$ 1.86 mins) &
1450.00\\
Fireman & 23743.36 & ($\approx$ 395.72 mins $\approx$ 6.6 hours) &
37.72 & ($\approx$ 0.62 mins) &
629.00 \\
Cooperative Navigation & 9741.37  & ($\approx$ 162 mins $\approx$ 2.7 hours) &
2.59 & ($\approx$ 0.04 mins) &
3754.00\\
\bottomrule
\end{tabular}
}
    \caption{Average running time in seconds of a run}
    \label{tab:runningtime}
\end{table}


The ratio of PyIBL running time to SpeedyIBL running time in Table \ref{tab:runningtime} shows that the benefit of SpeedyIBL over PyIBL increases significantly with the complexity of the task. In a simple task such as binary choice, SpeedyIBL performs 1.14 faster than PyIBL. However, the speed-up ratio increases with the higher dimensional state space tasks; for example, in Minimap SpeedyIBL was 279 times faster than PyIBL; and in Ms. Pac-Man SpeedyIBL was 1450 times faster than PyIBL.

Furthermore, the multi-agent tasks exhibit the largest ratio benefit of SpeedyIBL over PyIBL. For example, in the Cooperative Navigation task, PyIBL took about 2.7 hours to finish a run, but SpeedyIBL only takes 2.59 seconds to accomplish a run.

In all tasks, we observe that the computational time of SpeedyIBL is significantly shorter than running the same task in PyIBL; we also observe that there is no significant difference in the performance of SpeedyIBL and PyIBL ($p>0.05)$. These results suggest that SpeedyIBL is able to greatly reduce the execution time of an IBL model without compromising its performance. 

\begin{table}[!htbp]
    \centering
    \begin{tabular}{llrrr} 
\toprule
Task & Metric & PyIBL & SpeedyIBL & $t$-test\\
& & performance & performance &\\ 
\midrule
Binary choice& PMax & 0.833 & 0.828 & $t = -0.83, p = 0.4 > 0.05$\\
\hline 
Insider Attack Game & 
Average Reward & 1.383 & 1.375 & $t = -0.38, p= 0.69 > 0.05$ \\
Minimap & Average Reward & 
4.102 &
4.264 & $t = 0.87, p = 0.38>0.05$ \\
Ms.Pac-Man & Average Reward &
228.357 &
228.464 & $t = 0.72, p = 0.47 > 0.05$ \\
Fireman & Average Reward & 4.783 & 4.946 & $t = 1.07, p=0.28>0.05$ \\
Cooperative Navigation & Average Reward &
2.705 & 2.726 & $t = 0.69, p = 0.48 > 0.05$ \\
\bottomrule
\end{tabular}
    \caption{Average performance of a run of 100 episodes}
    \label{tab:reward}
\end{table}

\subsection{Learning curves}
\begin{figure}[!htbp]
\captionsetup{singlelinecheck = false, format= hang, justification=raggedright, font=footnotesize, labelsep=space}
\centering
\begin{subfigure}{\linewidth}
    \includegraphics[width=0.92\linewidth]{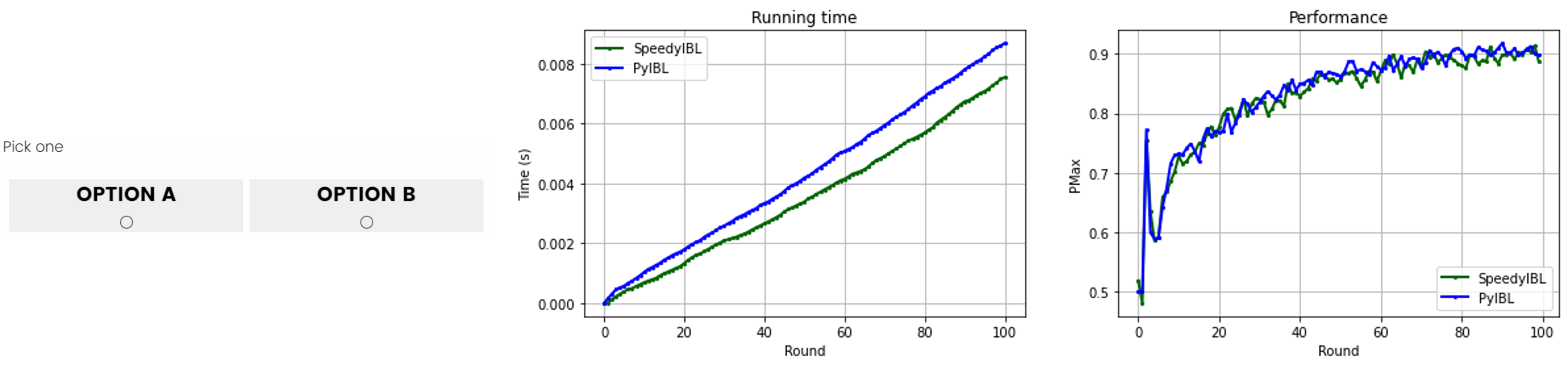}
    \caption{Binary Choice}
    \label{fig:binary-meta}
    \end{subfigure}
\begin{subfigure}{\linewidth}
    \includegraphics[width=0.92\linewidth]{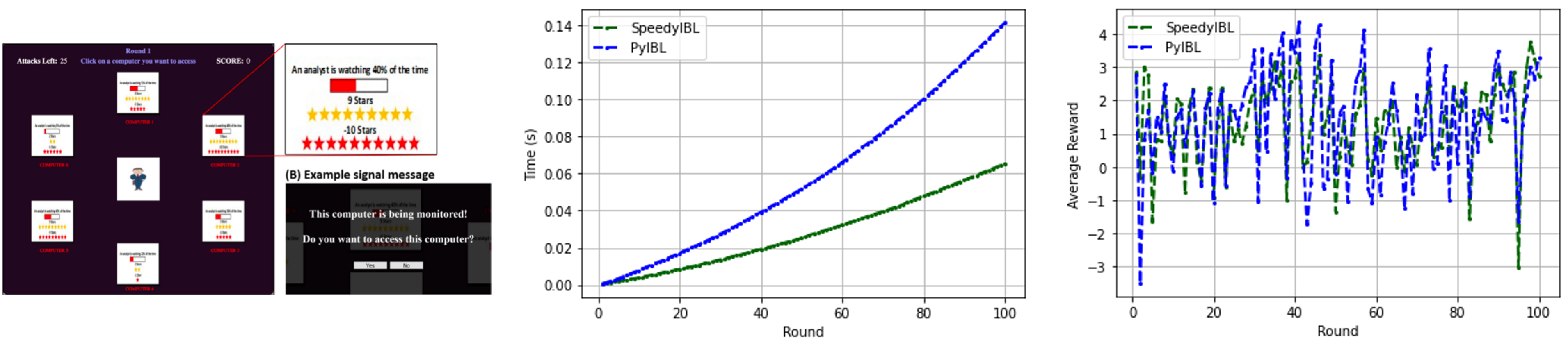}
    \caption{Insider Attack}
    \label{fig:insider-meta}
    \end{subfigure}
\begin{subfigure}{\linewidth}
    \includegraphics[width=0.92\linewidth]{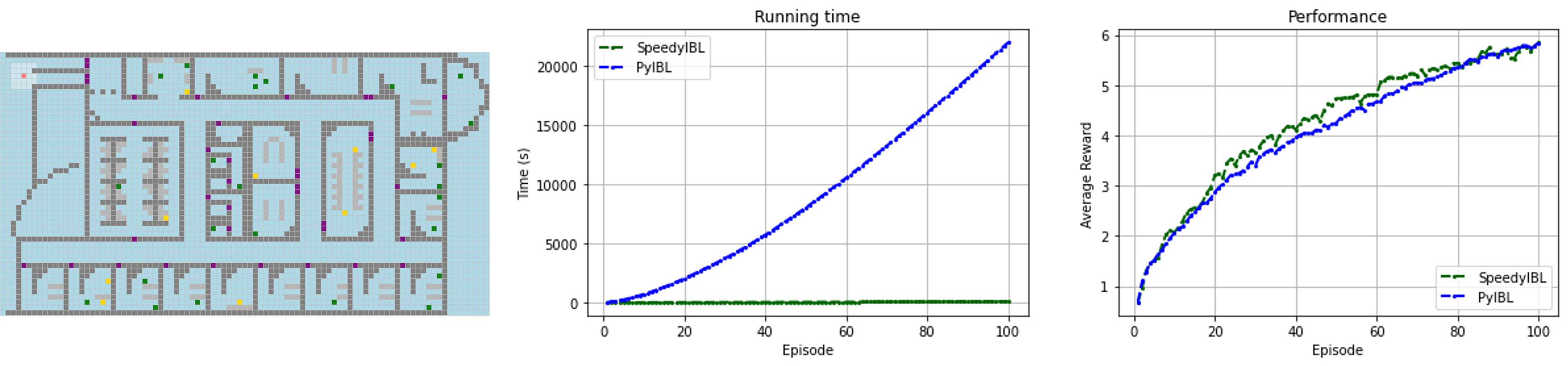}
    \caption{Minimap}
    \label{fig:minimap-meta}
    \end{subfigure}
\begin{subfigure}{\linewidth}
    \includegraphics[width=0.92\linewidth]{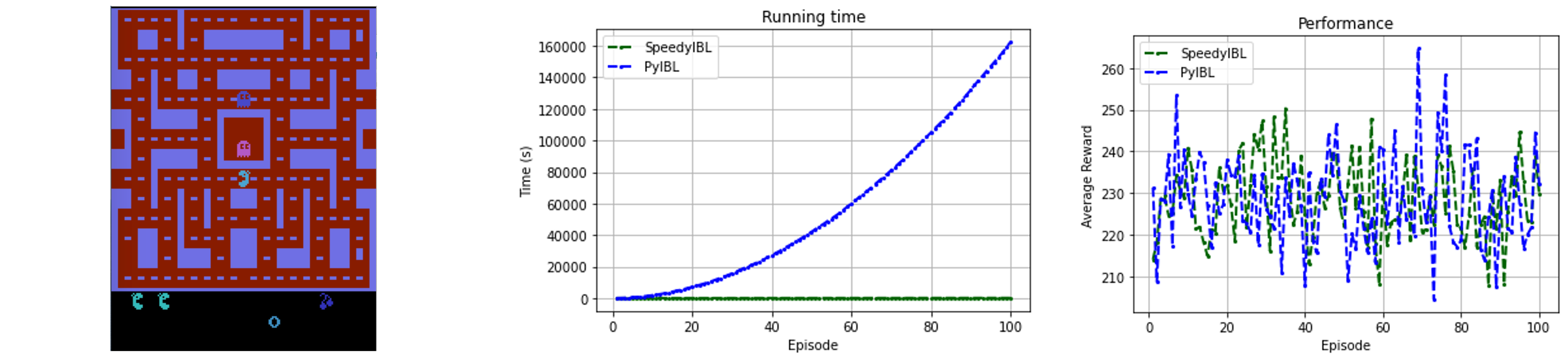}
    \caption{Ms.Pac-Man}
    \label{fig:mis-pacman-meta}
    \end{subfigure}
\begin{subfigure}{\linewidth}
    \includegraphics[width=0.92\linewidth]{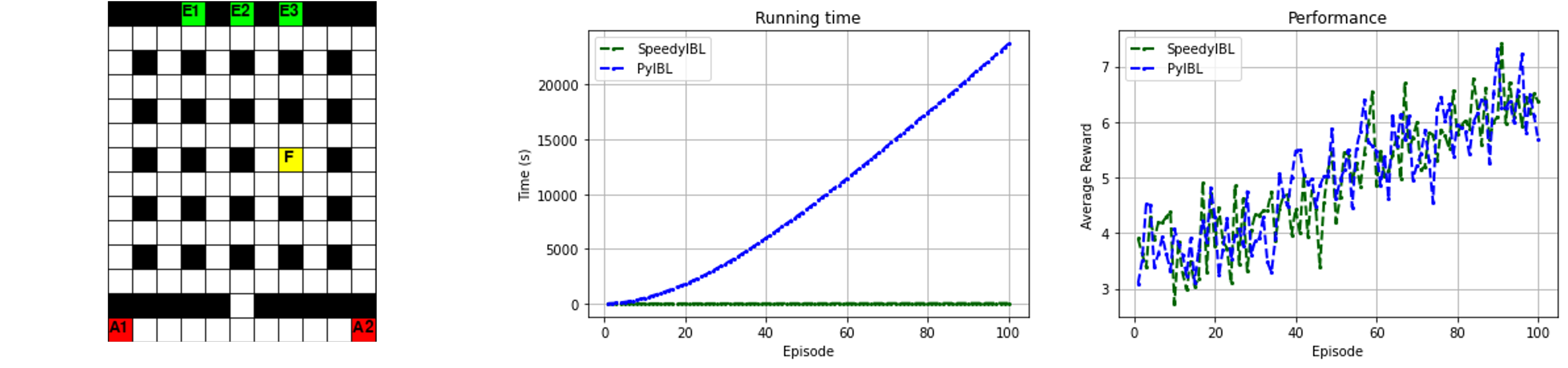}
    \caption{Fireman}
    \label{fig:fireman-meta}
\end{subfigure}
\begin{subfigure}{\linewidth}
    \includegraphics[width=0.92\linewidth]{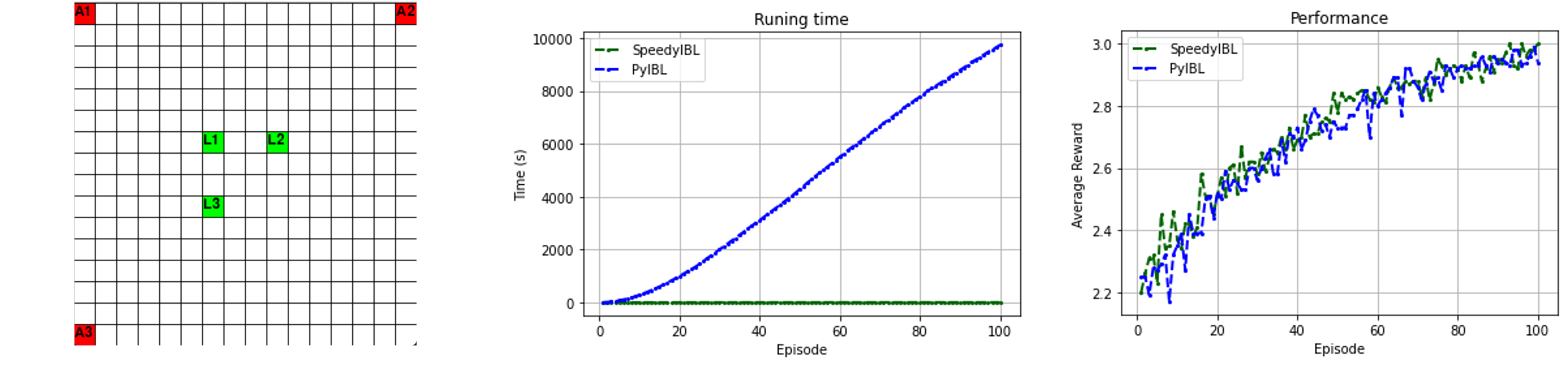}
    \caption{Cooperative Navigation}
    \label{fig:navigation-meta}
\end{subfigure}
\caption{The comparison between SpeedyIBL (Green line) and PyIBL (Blue line) over time in the considered tasks.}
\label{fig:comparison}
\end{figure}

Figure \ref{fig:comparison} shows the comparison of average running time (middle column) and average performance (right column) between PyIBL (Blue) and SpeedyIBL (Green) across episodes for all the six tasks.

\begin{figure}[!htbp]
    \centering
       \includegraphics[width=0.44\textwidth]{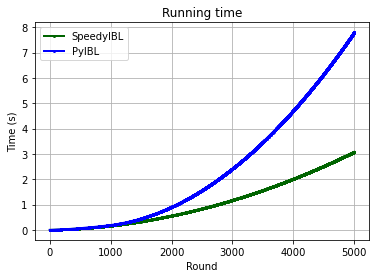}
    \includegraphics[width=0.45\textwidth]{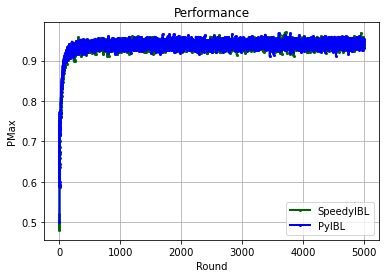}
    \caption{The comparison between SpeedyIBL and PyIBL in 5000 playing episodes of binary choice task.}
    \label{fig:binary_5000}
\end{figure}

In the \textit{Binary choice} task, it is observed that there is a small difference in the execution time before 100 episodes; where SpeedyIBL performs slightly faster than PyIBL. To illustrate how the benefit of SpeedyIBL over PyIBL implementation increases significantly as the number of episodes increase, we ran these models over 5000 episodes. The results in Figure~\ref{fig:binary_5000} illustrate the \textit{curse of exponential growth} very clearly, where PyIBL exponentially increases the execution time with more episodes. The benefit of SpeedyIBL over PyIBL implementation is clear with increased episodes. The PMax of SpeedyIBL and PyIBL overlap, again suggesting no different in their performance.

In the \textit{Insider Attack game} as shown Figure~\ref{fig:insider-meta}, the relation between SpeedyIBL and PyIBL in terms of computational time shows again, an increased benefit with increased number of episodes. We see that their running time is indistinguishable initially, but then the difference becomes distinct in the last 60 episodes. Regarding the performance (i.e., average reward), again, their performance over time is nearly identical. Learning in this task was more difficult, given the design of this task, and we do not observe a clear upward trend in the learning curve due to the presence of stochastic elements in the task. 

In all the rest of the tasks, the \textit{Minimap}, \textit{Ms.Pac-Man}, \textit{Fireman}, and \textit{Cooperative Navigation}, given the multi-dimensionality of these tasks representations and the number of agents involved in \textit{Fireman}, and \textit{Cooperative Navigation} tasks, the \textit{curse of exponential growth} is observed from early on, as shown in Figure~\ref{fig:minimap-meta}. The processing time of PyIBL grows nearly exponentially over time in all cases. The curve of SpeedyIBL also increases, but it appears to be constant in relation to the exponential growth of PyIBL given the significant difference between the two, when plotted in the same scale.  


The performance over time is again indistinguishable between PyIBL and SpeedyIBL.  Depending on the task, the dynamics, and strochastic elements of the task, the models' learning curves appear to fluctuate over time (e.g. \textit{Ms.Pac-Man}), but when the scenarios are consistent over time, the models show similar learning curves for both, PyIBL and SpeedyIBL. 



\section{Discussion and Conclusions}

Cognitive models are used increasingly to make predictions of human behavior and simulate the process by which humans make decisions from experience \cite{cranford2020toward,NGUYEN2020ICCM,Nguyen21}. In particular, many computational models have been developed relying on IBLT \cite{GONZALEZ03}. These IBL models have demonstrated how human decision processes are captured and characterized \cite{GONZALEZ11}, and most importantly, they provide evidence for the application and usefulness of the theory.

In this paper, we present an updated account of IBLT, the current formalization of its theoretical components and a comprehensive and precise presentations of the mechanisms of the theory. We aimed at improving the IBLT clarity and describing the mechanisms behind the general process of IBLT with precise mathematical representations and an algorithm implementation. Crucially, we demonstrated the generality and ability of the theory to predict human learning from experience in a wide variety of decision making tasks. That is, we provided a demonstration of how models grounded on the same IBLT can be applied and handle decision making tasks varying in the number of agents, the number of actions, the number of decision options and states, and the type of feedback delays.

We observed that implementing IBL models for these tasks using an existing library, PyIBL~\cite{MorrisonGonzalez}, comes at a practical cost. It is difficult to deal with the exponential growth of the memory of instances as more observations accumulate over time, which leads directly to an exponential slow down of the computational time when the characteristics of the tasks escalate from a single-agent to multi-agent and multi-state settings. Such problem is referred to as the curse of exponential growth, a common computational problem that emerges in many modeling approaches involving tabular computations. Clearly, resolving the curse of exponential growth becomes even more urgent when IBL models are expected to be increasingly used in interactive, real-time tasks that involve humans and models working together, similar to what has been shown recently in a number of RL initiatives \cite{carroll2019utility,strouse2021collaborating}.

To that end, we have developed a new implementation of IBL cognitive models called SpeedyIBL that not only employs a proper data structure for storing memory more efficiently, but also leverages the parallel computation using vectorization~\cite{larsen2000exploiting} to speed up the performance of IBL models in the presence of the curse of exponential growth.
We have assessed the robustness of SpeedyIBL by comparing it with PyIBL, a benchmark of the implementation of IBL models in Python~\cite{MorrisonGonzalez}, across a taxonomy of decision-making tasks varying in their increased complexity. We specifically demonstrated that SpeedyIBL implementation is able to perform considerably faster than PyIBL without compromising task performance. Moreover, the results also indicate that the difference in the running time of the SpeedyIBL and PyIBL becomes profound, especially in high-dimensional state spaces and multi-agent domains wherein more agents concurrently collaborate in a task.

Overall, we have introduced SpeedyIBL implementation that enables researchers to create multiple IBL agents relying on IBLT with fast processing and response time. SpeedyIBL can not only be used in simulation experiments of extended learning time, but also can be integrated into browser-based applications in which IBL agents can interact with human subjects in real-time. 
Given that the computation time of cognitive models in the literature is often overlooked, we believe that the techniques used in SpeedyIBL will be particularly useful for many other ACT-R cognitive models that are still built upon a heavyweight framework programmed in LISP. In that respect, numerous examples can be cited, including a cognitive multi-agent model \cite{REIhow2011}, a cognitive model for human-robot interaction \cite{LEBcog2013}, hybrid model consisting of a Deep RL agent and a cognitive model \cite{MITtow2021}, and many other models in the ACT-R literature\footnote{\url{http://act-r.psy.cmu.edu/publication/}}.
Moreover, provided that research on human–machine behavior has attracted much attention lately, we are convinced that SpeedyIBL will bring significant benefits to researchers and demonstrate the usefulness of IBL models in interactive tasks with human players.

\section*{Transparency and Openness}
SpeedyIBL is provided as a free and open-source Python library. All the codes, extensive documentation, simulation data, and all scripts used for analyses presented in this manuscript are available on Github \url{https://github.com/DDM-Lab/SpeedyIBL} and on OSF~\url{https://osf.io/gwqte/}. In addition, the Appendix provides a detailed tutorial including installation of the SpeedyIBL library and examples on how to replicate our demonstrations in the tasks offered in this paper.

\begin{acknowledgements}
This research was partly sponsored by the Defense Advanced Research Projects Agency and was accomplished under Grant Number W911NF-20-1-0006 and by AFRL Award FA8650-20-F-6212 subaward number 1990692 to Cleotilde Gonzalez.
\end{acknowledgements}

\clearpage
\appendix

\section*{Appendix: SpeedyIBL Tutorial} \label{sec:appendix}
In an attempt to increase the usage of SpeedyIBL, we hereby provide a tutorial on how to install and use the SpeedyIBL library, following exisiting research practice~\cite{evans2019method,henninger2021lab,vincent2016hierarchical}. Specifically, we explain how to build an IBL agent and elaborate on the meaning of associated inputs and functions. Afterwards, we present examples on two illustrative tasks:  Binary Choice~\ref{binarychoice} and Navigation~\ref{navigation}. It is worth noting that all the codes to run all the tasks and to reproduce the results presented in the paper are available at \url{https://github.com/DDM-Lab/SpeedyIBL}. In addition, we provide a Jupyter notebook file of the turorial, see \url{https://github.com/DDM-Lab/SpeedyIBL/blob/main/tutorial_speedyibl.ipynb}, for running the all tasks considered in this work using SpeedyIBL. We also make it available on Google Colab \url{https://colab.research.google.com/github/nhatpd/SpeedyIBL/blob/main/tutorial_speedyibl.ipynb}, where one can easily run it with no need to install Python and any relevant modules on their personal computers. 
Finally, we give a detailed instruction on how to reproduce all the reported results using PyIBL and SpeedyIBL. 

\subsection*{Installing SpeedyIBL}
Note that the SpeedyIBL library is a Python module, which is stored at PyPI (\url{pypi.org}), a repository of software for the Python programming language, see \url{https://pypi.org/project/speedyibl/}. Hence, installing SpeedyIBL is a very simple process. Indeed, one can install SpeedyIBL by simply typing the following line in a command prompt:

\begin{lstlisting}[language=Python]
pip install speedyibl
\end{lstlisting}
\subsection*{Describing an Agent with SpeedyIBL}
After installing the library, we need to import the class \texttt{Agent} of SpeedyIBL by typing:

\begin{lstlisting}[language=Python]
from speedyibl import Agent
\end{lstlisting}
We provide the descriptions of the inputs and main functions of the class \texttt{Agent} in the following tables. \\
\\
\resizebox{.99\textwidth}{!}{%
    \begin{tabular}{lll} 
\toprule
Inputs & Type & Description\\
\hline
default\_utility & float or None & initial utility value for each instance, default = 0.1\\
& & or None if prepopulated \\
noise & float & noise parameter $\sigma$, default  = 0.25\\
decay & float &  decay paremeter $d$, default = 0.5\\
mismatchPenalty & float or None & mismatch penalty parameter, default = None (without partial matching process)\\
lendeque & int or None & maximum size of a deque for each instance that contains  \\
& & timestamps or  None if unbounded, default = 250000\\
\bottomrule
\end{tabular}
}
\\


\noindent \resizebox{.99\textwidth}{!}{%
    \begin{tabular}{lll} 
\toprule
Functions & Inputs & Description\\
\hline
choose & list of options & choose one option from the given list of options\\
respond & reward & add the current timestamp to the instance \\
& & of the last option and reward\\ 
prepopulate & option, reward & initialize time 0 for the instance of this option and reward\\
populate\_at & option, reward, time & add time to the instance of this option and reward\\
equal\_delay\_feedback & reward, list of instances,& update instances in the list by using this reward\\
instances & no input & show all the instances in the memory\\
\bottomrule
\end{tabular}
}

\subsection*{Using SpeedyIBL for Binary Choice Task}
From the list of inputs of the class \texttt{Agent}, although we need five inputs to create an IBL agent, by using the defaults for noise, decay, mismatchPenalty, and lendeque, we only need to pass the value of \texttt{default\_utility} (here in the example is 4.4). Hence we create an IBL agent for the binary choice task as follows:
\begin{lstlisting}[language=Python]
agent = Agent(default_utility=4.4)
\end{lstlisting}
We then define a list of options for the agent to choose:
\begin{lstlisting}[language=Python, upquote=true]
options = ['A','B'] # A is the safe option while B is the risky one
\end{lstlisting}
We are now ready to make the agent \texttt{choose} one of the two options:
\begin{lstlisting}[language=Python]
choice = agent.choose(options)
\end{lstlisting}
Next, we determine a reward that the agent can receive after choosing one of the options, see Subsection \ref{binarychoice}:
\begin{lstlisting}[language=Python, upquote=true]
import random 
if choice == 'A':
    reward = 3
elif random.random() <= 0.8:
    reward = 4
else:
    reward = 0
\end{lstlisting}
After choosing one option and observing the reward, we use the function \texttt{respond}, see the table above, to store the instance in the memory as follows:
\begin{lstlisting}[language=Python]
agent.respond(reward)
\end{lstlisting}
That is, we have run one trial for the binary choice task, which the process includes choosing one option, observing the reward, and storing the instance (\texttt{respond}). To conduct 1000 runs of 100 trials, we use two for loops as follows:
\begin{lstlisting}[language=Python, upquote=true]
import time # to calculate time
runs = 1000 # number of runs (participants)
trials = 100 # number of trials (episodes)
average_p = [] # to store average of performance (proportion of maximum reward expectation choice)
average_time = [] # to save time 
for r in range(runs):
  pmax = []
  ttime = [0]
  agent.reset() #clear the memory for a new run
  for i in range(trials):     
    start = time.time()
    choice = agent.choose(options) # choose one option from the list of two
    # determine the reward that agent can receive
    if choice == 'A':
        reward = 3
    elif random.random() <= 0.8:
        reward = 4
    else:
        reward = 0
    # store the instance
    agent.respond(reward)
    end = time.time()
    ttime.append(ttime[-1]+ end - start)
    pmax.append(choice == 'B') 
  average_p.append(pmax) # save performance of each run 
  average_time.append(ttime) # save time of each run 
\end{lstlisting}
Finally, we provide the following code to plot the running time and performance of this SpeedyIBL agent.

\begin{lstlisting}[language=Python, upquote=true]
import matplotlib.pyplot as plt
import numpy as np 
plt.rcParams["figure.figsize"] = (12,4)
plt.subplot(int('12'+str(1)))
plt.plot(range(trials+1), np.mean(np.asarray(average_time),axis=0), 'o-', color='darkgreen', markersize=2, linestyle='--', label='speedyIBL')
plt.xlabel('Round')
plt.ylabel('Time (s)')
plt.title('Runing time')
plt.legend()
plt.subplot(int('12'+str(2)))
plt.plot(range(trials), np.mean(np.asarray(average_p),axis=0), 'o-', color='darkgreen', markersize=2, linestyle='--', label='speedyIBL')
plt.xlabel('Round')
plt.ylabel('PMAX')
plt.title('Performance')
plt.legend()
plt.grid(True)
plt.show()
\end{lstlisting}
It is worth noting that the codes of both SpeedyIBL and PyIBL for generating the results of the binary choice task in the paper are available at \url{https://github.com/DDM-Lab/SpeedyIBL/blob/main/Codes/binarychoice.py}. To plot the results, please see \url{https://github.com/DDM-Lab/SpeedyIBL/blob/main/Codes/plot_results.ipynb}. 

\subsection*{Using SpeedyIBL for Cooperative Navigation task}
First, let us build an environment class of the cooperative navigation task. Although constructing an environment depends on specific tasks, it consists of two main functions: \texttt{reset} and \texttt{step}. The \texttt{reset} function sets the agents to their starting locations at beginning of each episode while the \texttt{step} function moves the agents to new locations and returns a new state, reward, and task status (task finished or not) after they made decisions. 

\noindent We would like to note that we created a Python module \texttt{vitenv} containing all the environments of the tasks considered in the paper, which can be accessed at \url{https://pypi.org/project/vitenv/}. The codes of the environments of other tasks and this tutorial also available at our Github link \url{https://github.com/DDM-Lab/SpeedyIBL}. Below is an illustrative code of building the environment of the cooperative navigation task:
\begin{lstlisting}[language=Python, upquote=true]
import numpy as np
import copy
class Environment(object):
  def __init__(self):
    #Initialize elements if the task including size of grid-world, number of agents, pixel values of agents, landmark, initial locations of agents and landmarks 
    self.GH = 16 #height of grid world
    self.GW = 16 #width of grid world
    self.NUMBER_OF_AGENTS = 3 #number of agents
    self.AGENTS = [240.0, 200.0, 150] #pixel values of [Agent1, Agent2, Agent3]
    self.LANDMARKS = [40, 50, 60] #pixel values of landmarks
    self.AGENTS_X = [0, self.GW-1, 0] #x-coordinates of initial locations of agents
    self.AGENTS_Y = [0, 0, self.GH-1]  #y-coordinates of initial locations of agents
    MID = 8
    self.LANDMARK_LOCATIONS = [(MID-2,MID-2),(MID+1,MID-2),(MID-2, MID+1)] #locations of landmarks
    self.ACTIONS = 4 # move up, down, left, righ
    
  def reset(self):
    # Reset everything. 
    self.s_t = np.zeros([self.GH,self.GW], dtype=np.float64) #create an array that represents states of the grid-world 
    # Agents and landmarks are initialised:
    # Agent x and y positions can be set in the following lists.
    self.agents_x = copy.deepcopy(self.AGENTS_X)
    self.agents_y = copy.deepcopy(self.AGENTS_Y)
    self.agent_status= [False for i in range(self.NUMBER_OF_AGENTS)]
    #Set pixel values of agents
    for i in range(self.NUMBER_OF_AGENTS):
        self.s_t[self.agents_y[i]][self.agents_x[i]] += self.AGENTS[i]
    #Initialize the landmarks in the environment
    for l,p in zip(self.LANDMARK_LOCATIONS,self.LANDMARKS):
      self.s_t[l[1]][l[0]] = p
    self.reached_landmarks = []
    return self.s_t
  def step(self, actions):
  # Change environment state based on actions. :param actions: List of integers providing actions for each agent
    #Move agents according to actions.
    for i in range(self.NUMBER_OF_AGENTS):
      if not self.agent_status[i]:
        dx, dy = self.getDelta(actions[i])
        targetx = self.agents_x[i] + dx
        targety = self.agents_y[i] + dy
        if self.noCollision(targetx, targety):
          self.s_t[self.agents_y[i]][self.agents_x[i]] -= self.AGENTS[i]
          self.s_t[targety][targetx] += self.AGENTS[i]
          self.agents_x[i] = targetx
          self.agents_y[i] = targety
          if (targetx,targety) in self.LANDMARK_LOCATIONS:
            self.agent_status[i] = True
            if (targetx,targety) not in self.reached_landmarks:
              self.reached_landmarks.append((targetx,targety))
    terminal = sum(self.agent_status) == 3
    reward = len(self.reached_landmarks)
    return self.s_t, reward, terminal 
  def getDelta(self, action):
    # Determine the direction that the agent should take based upon the action selected. The actions are: 'Up':0, 'Right':1, 'Down':2, 'Left':3, :param action: int
    if action == 0:
      return 0, -1
    elif action == 1:
      return 1, 0    
    elif action == 2:
      return 0, 1    
    elif action == 3:
      return -1, 0
    elif action == 4:
      return 0, 0 
  def noCollision(self, x, y):
    # Checks if x, y coordinate is currently empty :param x: Int, x coordinate :param y: Int, y coordinate
    if x < 0 or x >= self.GW or\
      y < 0 or y >= self.GH or\
      self.s_t[y][x] in self.AGENTS:
      return False
    else:
      return True
\end{lstlisting}
Now, we can call the environment and \texttt{reset} it as follows:
\begin{lstlisting}[language=Python, upquote=true]
env = Environment()
s = env.reset()
\end{lstlisting}
Like in the binary choice task, we define three agents with \texttt{default\_utility}=2.5 and save them in a list agents:
\begin{lstlisting}[language=Python, upquote=true]
number_agents = 3
agents = []
episode_history = {}
for i in range(number_agents): 
  agents.append(Agent(default_utility=2.5))
  episode_history[i] = []
\end{lstlisting}
Here we have used a dictionary \texttt{episode\_history} to save information of each episode that we will use for the delay feedback mechanism.  Next, we create a list of options:
\begin{lstlisting}[language=Python, upquote=true]
s_hash = hash(s.tobytes())
options = [(s_hash, a) for a in range(env.ACTIONS)]
\end{lstlisting}
Here we have used the \texttt{hash} function to convert an array into a hashable object used as a key in a Python dictionary. Now we make the agents \texttt{choose} their options and save instances.
\begin{lstlisting}[language=Python, upquote=true]
actions = [4,4,4]
for i in range(number_agents):
  if not env.agent_status[i]:
    option = agents[i].choose(options)
    actions[i] = option[1]
    agents[i].respond(0)
    episode_history[i].append((option[0],option[1],0,agents[i].t))
\end{lstlisting}
After choosing actions, the locations of the agents are updated in the environment by the \texttt{step} function:
\begin{lstlisting}[language=Python, upquote=true]
s, reward, t = env.step(actions)
\end{lstlisting}
When the agents finish the task (reach landmarks, i.e., t = True) or when they reach the maximum number of steps, we update outcomes of previous instances by an equal delayed feedback mechanism.
\begin{lstlisting}[language=Python, upquote=true]
for i in range(number_agents):
    agents[i].equal_delay_feedback(reward, episode_history[i])
\end{lstlisting}
In order to run 100 times of 100 episodes with 2500 steps, we use the code below.
\begin{lstlisting}[language=Python, upquote=true]
import time # to calculate time
runs = 100 # number of runs (participants)
trials = 100 # number of trials (episodes)
steps = 2500 # number of steps 
average_p = [] # to store average of performance (proportion of maximum reward expectation choice)
average_time = [] # to save time 

for r in range(runs):
  preward = []
  ttime = [0]
  #clear the memory for a new run
  for i in range(number_agents): 
    agents[i].reset()
  episode_history[i] = []
  for e in range(trials):     
    start = time.time()
    s = env.reset()
    # clear the previous episode
    for i in range(number_agents): 
      episode_history[i] = []
    for j in range(steps):
      s_hash = hash(s.tobytes())
      options = [(s_hash, a) for a in range(env.ACTIONS)]
      # choose one option from the list
      actions = [4,4,4]
      for i in range(number_agents):
        if not env.agent_status[i]:
          option = agents[i].choose(options)
          actions[i] = option[1]
          agents[i].respond(0)
          episode_history[i].append((option[0],option[1],0,agents[i].t)) # save information
      s, reward, t = env.step(actions)
      if t or j == steps-1:
        for i in range(number_agents):
          agents[i].equal_delay_feedback(reward, episode_history[i])
        break 
    end = time.time()
    ttime.append(ttime[-1]+ end - start)
    preward.append(reward) # save reward of each episode
  average_p.append(preward) # save performance of each run 
  average_time.append(ttime) # save time of each run 
\end{lstlisting}
To plot the results of the task, we can use the same source code as provided in the binary choice task.

\subsection*{Reproducing Results}
All the results can be reproduced by running corresponding scripts for each task under folder \texttt{\textbf{Codes}}. In particular, to run the tasks with SpeedyIBL or PyIBL, one can simply execute the following commands and the experiment will start. \\

\noindent 1. \textbf{Binary Choice Task}: 
\begin{lstlisting}[language=Python]
python3 binarychoice.py --method [name]
\end{lstlisting}
With argument \texttt{[name]} is replaced by: \texttt{libl} for SpeedyIBL and \texttt{ibl} for PyIBL. \\

\noindent 2. \textbf{Insider Attack Game}: 
\begin{lstlisting}[language=bash]
    python3 insider_attack_speedyIBL.py # to run SpeedyIBL
    python3 insider.py # to run PyIBL
\end{lstlisting} 

\noindent 3. \textbf{Minimap}: 
\begin{lstlisting}[language=Python]
python3 minimap.py --type [name]
\end{lstlisting}
With argument \texttt{[name]} is replaced by: \texttt{libl} for SpeedyIBL and \texttt{ibl} for PyIBL. \\

\noindent 4. \textbf{MisPac-man}: 
\begin{lstlisting}[language=Python]
python3 mispacman.py --type [name]
\end{lstlisting}
With argument \texttt{[name]} is replaced by: \texttt{libl} for SpeedyIBL and \texttt{ibl} for PyIBL. \\

\noindent 5. \textbf{Fireman}: 
\begin{lstlisting}[language=Python]
python3 fireman.py --type [name]
\end{lstlisting}
With argument \texttt{[name]} is replaced by: \texttt{libl} for SpeedyIBL and \texttt{ibl} for PyIBL. \\

\noindent 6. \textbf{Cooperative Navigation}: 
\begin{lstlisting}[language=Python]
python3 navigation.py --type [name]
\end{lstlisting}
With argument \texttt{[name]} is replaced by: \texttt{libl} for SpeedyIBL and \texttt{ibl} for PyIBL. \\


%
%

\clearpage
\bibliographystyle{spmpsci}      

%
%

%
\end{document}